\documentclass[12pt]{iopart}
\usepackage{iopams}
\usepackage{amssymb,latexsym}
\usepackage{amstext}

  \newcommand{\ket}[1]{\vert\mathit{#1}\rangle}
\newcommand{\bra}[1]{\langle\mathit{#1}\vert}
\newcommand{\braket}[2]{\langle\mathit{#1}\vert\mathit{#2}\rangle}

\begin{document}

\title[Feynman's path integral and MUBs]{Feynman's
       path integral and mutually unbiased bases}
 \author{J Tolar and G Chadzitaskos}
\address{Department of Physics\\
Faculty of Nuclear Sciences and Physical Engineering         \\
Czech Technical University in Prague\\ B\v rehov\'a 7,  CZ - 115 19
Prague, Czech Republic}
 \ead{jiri.tolar@fjfi.cvut.cz, goce.chadzitaskos@fjfi.cvut.cz}

 \begin{abstract}
Our previous work on quantum mechanics in Hilbert spaces of finite
dimensions $N$ is applied to elucidate the deep meaning of Feynman's
path integral pointed out by G. Svetlichny. He speculated that the
secret of the Feynman path integral may lie in the property of
mutual unbiasedness of temporally proximal bases. We confirm the
corresponding property of the short-time propagator by using a
specially devised $N \times N$-approximation of quantum mechanics in
$L^2 (\mathbb{R})$ applied to our finite-dimensional analogue of a
free quantum particle.
\end{abstract}

 \submitto{J. Phys. A: Math. Theor.}



\section{Introduction}

The circle of ideas which generated Feynman path integrals is
contained in works by Dirac and Feynman \cite{Feyn,FeynHibbs}.
Especially in the latter work the representation of quantum
mechanical evolution amplitudes in terms of heuristic integrals on
'path space' was developed into an alternative general formulation
of quantum dynamics, equivalent to the previous formulations by
Heisenberg, Schr\"odinger or Schwinger.

The miraculous success of Feynman's method in dealing with quantum
fields should be, however, mitigated by the fact that there is so
far no rigorous way to define it in terms of conventional measure
theory. Namely, the heuristic expression for Feynman path integrals
is in terms of a complex formal density which does not define a
measure. Thus the mathematical definition of the objects understood
under the name of Feynman path integrals posed genuine new problems
which have been attacked by different methods. For instance,
Feynman's path integral in non-relativistic quantum mechanics is
conventionally viewed as a formal expression which can be given
meaning by a specially devised limiting process.

The diverse aspects and approaches make clear that the subject of
Feynman path integrals should not be considered as a closed one, on
the contrary, much work is needed on the conceptual mathematical and
physical level in order to bring to fruition all the beautiful
potentialities contained in those ideas. In this direction it was
recently speculated by G. Svetlichny
\cite{Svetlichny07a,Svetlichny07b} that the secret of the path
integral may rest on the mutual unbiasedness of temporally proximal
bases. He focused the essential problems into the following
questions:
\begin{enumerate}
\item For what unitary groups $U(t)$ in $L^2(\mathbf{R}^n)$ do the
position bases at times $0$ and $t$ tend to mutual unbiasedness as
$t\rightarrow 0$ ?
\item Is there a discrete version of the previous question in a
finite-dimensional Hilbert space which approximately simulates the
propagation of a free particle?
\item What is the information-theoretic nature of the normalization
factor $A$ in the short-time propagator?
\end{enumerate}
In this paper we elucidate the true meaning of these questions in
the case of a \textit{non-relativistic quantum particle on the real
line}.

In section 2 we recall related notions of complementary observables
and of mutually unbiased bases. Then the basic concepts of quantum
mechanics in finite-dimensional Hilbert spaces are introduced in
section 3. Section 4 is devoted to the construction of finite
quantum phase space from the finite Heisenberg group. The group of
inner automorphisms of finite quantum phase space is described in
section 5. After these prerequisites, the $N \times N$-approximation
of quantum mechanics on the real line is constructed in section 6
and applied to an analogue of a free non-relativistic particle. This
approximation, being used in the Feynman short-time propagator
(section 7) leads to the emergence of a Lagrangian as the
corresponding local phase and at the same time demonstrates mutual
unbiasedness of temporally proximal bases (section 8). Our
derivation also yields the normalization factor $A$ as direct
counterpart of the constant $1/\sqrt{N}$ involved in the definition
of mutual unbiasedness.

\section{Complementarity and mutually unbiased bases}

Mutually unbiased bases in Hilbert spaces of finite dimensions are
closely related to the quantal notion of complementarity. Namely,
two observables $A$ and $B$ of a quantum system with Hilbert space
of finite dimension $N$ are called \textit{complementary}
\cite{Kraus}, if their eigenvalues are non-degenerate and any two
normalized eigenvectors $\ket{u_i}$ of $A$ and $\ket{v_j}$ of $B$
satisfy
     \begin{equation*}
     |\braket{u_i}{v_j}| = \frac{1}{\sqrt{N}}.
    \end{equation*}
Then in an eigenstate $\ket{u_i}$ of $A$ all eigenvalues
$b_1,\ldots,b_N$ of $B$ are measured with equal probabilities, and
vice versa. This means that exact knowledge of the measured value of
$A$ implies \textit{maximal uncertainty} to any measured value of
$B$.

According to W.K. Wootters \cite{Woottersprob}, two orthonormal
bases in an $N$-dimensional complex Hilbert space
 \begin{equation*}
 \lbrace \ket{u_i}
|i=1,2,\ldots,N \rbrace \, \mbox{ and } \, \left\lbrace\ket{v_j}
|j=1,2,\ldots,N\right\rbrace
 \end{equation*}
are called \textit{mutually unbiased}, if inner products between all
possible pairs of vectors taken from distinct bases have the same
magnitude $ 1 / \sqrt{N} $,
     \begin{equation*}
     |\braket{u_i}{v_j}| = \frac{1}{\sqrt{N}} \,
\mbox{ for all } \, i,j \in \left\lbrace 1,2,\ldots,N \right\rbrace.
    \end{equation*}
Thus if the system is in the state $\ket{u_i}$, then transitions to
any of the states $\ket{v_j}$ have equal probabilities.

It is important to note that, in an $N$-dimensional Hilbert space,
there cannot be more than $N+1$ mutually unbiased bases. It has also
been proved that the maximal number of $N+1$ mutually unbiased bases
is attained, if $N$ is a power of a prime \cite{WoottersFields}.

\section{Quantum mechanics in finite-dimensional Hilbert spaces}

The mathematical arena for ordinary quantum mechanics is, due to
Heisenberg's commutation relations, the infinite-dimensional Hilbert
space. A useful model for quantum mechanics in a Hilbert space of
finite dimension $N$ is due to H. Weyl \cite{Weyl}. Its geometric
interpretation as the simplest quantum kinematic on a finite
discrete configuration space formed by a periodic chain of $N$
points was elaborated by J.~Schwinger \cite{Schwinger*}. In
\cite{StovTolar} we proposed a group theoretical formulation of this
quantum model as well as a finite-dimensional analogue of quantum
evolution operator for a free particle.

In an $N$-dimensional Hilbert space with orthonormal basis
$\mathcal{B} = \left\lbrace\ket{0}, \ket{1}, \ldots
\ket{N-1}\right\rbrace$ the Weyl pair of unitary operators $(Q_N,
P_N)$ is defined by the relations
    \begin{eqnarray*}
 Q_N \ket{\rho} = \omega_N^\rho \ket{\rho}, \quad \rho=0,1,\ldots,N-1, \\
 P_N \ket{\rho} = \ket{\rho-1 \pmod{N}},
    \end{eqnarray*}
where $\omega_N = \exp(2\pi i/N)$ \cite{Weyl} (see also
\cite{Vourdas,Kibler}). If $\mathcal{B}$ is the canonical basis of
$\mathbb{C}^N$, the operators $Q_N$ and $P_N$ are represented by the
matrices
    \begin{equation*}
 Q_N = \mbox{diag}\left(1,\omega_N,\omega_N^2,\cdots,\omega_N^{N-1}\right)
    \end{equation*}
and
    \begin{equation*}
     P_N = \left(
    \begin{array}{cccccc}
     0 & 1 & 0& \cdots & 0 & 0 \\
     0 & 0& 1&  \cdots & 0 & 0 \\
     0 & 0 & 0&\cdots & 0 & 0 \\
     \vdots &   & & \ddots &   & \\
     0 & 0 &0 & \cdots & 0 & 1 \\
     1 & 0 &0 &\cdots & 0 & 0
    \end{array} \right)
    \end{equation*}
They fulfil a commutation relation
    \begin{equation}
     \label{qnpnomega}
    P_N Q_N =   \omega_N Q_N P_N
    \end{equation}
which is analogous to the relation for Weyl's exponential form of
Heisenberg's commutation relations. Further, $P_N^N = Q_N^N = I_N$,
$\omega_N^N = 1$.

The \textit{finite Heisenberg group} is generated by $\omega_N$,
$Q_N$ and $P_N$
     \begin{equation*}
     \Pi_N = \left\lbrace \omega_N^l Q^j_N P_N^\sigma |
     l,j,\sigma = 0,1,2,\ldots,N-1\right\rbrace.
     \end{equation*}
It consists of $N^3$ unitary $N \times N$ matrices and is also
called the Pauli group.

The \textit{geometrical picture} behind the above operators is the
following \cite{StovTolar}. The cyclic group $
 \mathbb{Z}_N = \left\lbrace 0,1,\ldots N-1
 \right\rbrace $
is the configuration space for $N$-dimensional quantum mechanics.
Elements of this periodic chain $\mathbb{Z}_N$ provide labels of the
vectors of the basis $\mathcal{B} = \left\lbrace\ket{0}, \ket{1},
\ldots \ket{N-1}\right\rbrace$ with the physical interpretation that
$\ket{\rho}$ is the (normalized) eigenvector of position at $\rho
\in \mathbb{Z}_N$. The action of $\mathbb{Z}_N $ on $\mathbb{Z}_N $
via addition modulo $N$ is represented by unitary operators
$U(\sigma) = P_{N}^{\sigma}$. The action of these discrete
translations on vectors $\ket{\rho}$ from basis $\mathcal{B}$ is
given by
  \begin{equation*}
    U(\sigma)\ket{\rho} = P_N^\sigma \ket{\rho}=\ket{\rho-\sigma \pmod{N}}.
  \end{equation*}

The important \textit{discrete Fourier transformation} is given by
the unitary Sylvester matrix $S_N$ with elements
   \begin{equation*}
(S_N)_{k \rho}=\braket{\rho}{k}= \frac{\omega_N^{\rho k}}{\sqrt{N}}.
   \end{equation*}
The operator relation
   \begin{equation*}
   \label{sylvesterPSylevster}
    S_N^{-1} P_N S_N = Q_N
   \end{equation*}
shows that the discrete Fourier transform diagonalizes the momentum
operator. In other words, it performs the transition from the
coordinate representation to the momentum representation:
  \begin{equation*}
  \ket{k}= \sum_{\rho=0}^{N-1}\ket{\rho}\braket{\rho}{k}.
\end{equation*}

\section{Finite quantum phase space}
The following developments will heavily use the \textit{finite phase
space} $\Gamma_N$ which is simply related to the finite Heisenberg
group \cite{Balian}. The center $Z(\Pi_N)$ of the finite Heisenberg
group is the set of all those elements of $\Pi_N$ which commute with
all elements in $\Pi_N$
    \begin{equation*}
Z(\Pi_N) = \left\lbrace (l,0,0) |l=0,1,\ldots,N-1\right\rbrace.
    \end{equation*}
Since the center is a normal subgroup, one can go over to the
quotient group $\Pi_N / Z(\Pi_N)$. Its elements are the cosets
labeled by pairs $(j,\sigma)$, $j,\sigma = 0,1,\ldots,N-1$. The
quotient group can be identified with the finite phase space
 $$\Gamma_N =
\mathbb{Z}_N \times \mathbb{Z}_N, \quad N=2,3,\dots.
 $$
To simplify notation, we denote the cosets corresponding to elements
$(j,\sigma)$ of the phase space $ \Gamma_N$ by $Q^j P^\sigma$
without subscripts $N$
    \begin{equation*}
Q^j P^\sigma= \left\lbrace \omega_N^l Q^j_N P_N^\sigma | \quad l =
0,1,\ldots,N-1 \right\rbrace.
    \end{equation*}
The correspondence
    \begin{equation*}
\phi : \Pi_N / Z(\Pi_N) \rightarrow \Gamma_N=\mathbb{Z}_N \times
\mathbb{Z}_N  :  Q^j P^\sigma \mapsto (j,\sigma),
     \end{equation*}
     is an isomorphism of Abelian groups, since
     \begin{equation*}
 \phi\left(\left(Q^j P^\sigma\right)\left(Q^{j'}
P^{\sigma'}\right)\right) = \phi\left(\left(Q^j
P^\sigma\right)\right)\phi\left(\left(Q^{j'}
P^{\sigma'}\right)\right) =
\end{equation*}
 \begin{equation*}
 = (j,\sigma) + (j',\sigma') = (j+j',\sigma+\sigma').
     \end{equation*}

The group of automorphisms of the quantum phase space $\Gamma_N$ was
studied in \cite{Balian,Pateraspol}. The latter paper
considered---instead of cosets---the one-dimensional grading
subspaces of the Pauli-graded Lie algebra $gl(N,\mathbb{C})$ and
studied their transformations under the automorphisms of
$gl(N,\mathbb{C})$. The subgroup of \textit{inner automorphisms} was
induced by the action
     \begin{equation*}
      \psi_X (A) = X^{-1} A X
    \end{equation*}
of matrices $X$ from $GL(N,\mathbb{C})$.

Along the same vein we are considering those automorphisms of the
above form, acting on elements of $\Pi_N$, which induce permutations
of cosets in $\Pi_N / Z(\Pi_N)$. Operators $X$ which induce these
automorphisms are unitary. Explicit forms of these operators are
given in \cite{Balian} for $N$ prime and in \cite{Pateraspol} for
arbitrary $N$ but only for special transformations of $\Gamma_N$.

Automorphisms $\psi$ of the given form are equivalent if they define
the same transformation of cosets in $\Pi_N / Z(\Pi_N)$:
    \begin{equation*}
\psi_Y \sim \psi_X \quad \Leftrightarrow \quad
 Y^{-1} Q^j P^\sigma Y = X^{-1} Q^j P^\sigma X
    \end{equation*}
for all $(i,j) \in \mathbb{Z}_N \times \mathbb{Z}_N$. The group
$\Pi_N / Z(\Pi_N)$ has two generators, the cosets $P$ and $Q$. Hence
if $\psi_Y$ induces a transformation of $\Pi_N / Z(\Pi_N)$, then
there must exist elements $a,b,c,d \in \mathbb{Z}_N$ such that
     \begin{equation*}
  Y^{-1} Q Y = Q^a P^b \quad \mbox{and} \quad Y^{-1} P Y = Q^c P^d.
     \end{equation*}
It follows that to each equivalence class of automorphisms $\psi_Y$
a quadruple $(a,b,c,d)$ of elements in $\mathbb{Z}_N$ is assigned.
Then we have

  {\bf Theorem} \cite{Pateraspol} \textit{There is an
isomorphism $\Phi$ between the set of equivalence classes of inner
automorphisms $\psi_Y$ and the group $SL(2,\mathbb{Z}_N)$ of $2
\times 2$ matrices with determinant equal to $1 \mbox{ modulo } N$,
       \begin{equation*}
        \Phi(\psi_Y) = \left( \begin{array}{cc}
        a & b \\
        c & d
       \end{array}\right), \qquad a,b,c,d \in \mathbb{Z}_N;
       \end{equation*}
the action of these automorphisms on $\Pi_N / Z(\Pi_N)$ is given by
the right action of $SL(2,\mathbb{Z}_N)$ on the phase space
$\Gamma_N=\mathbb{Z}_N \times \mathbb{Z}_N$,
       \begin{equation*}
        (j',\sigma') =  (j,\sigma) \left( \begin{array}{cc}
        a & b \\
        c & d
       \end{array}  \right).
       \end{equation*}
}

\section{$N\times N$ approximation of quantum mechanics
       on the real line}

An interesting approximation method in quantum mechanics was
proposed in E. Husstad's PhD. thesis \cite{Husstad} supervised by T.
Digernes at NTNU Trondheim. Their approach was inspired by an idea
of J. Schwinger \cite{Schwinger**}.

They approximate quantum operators in $L^2(\mathbb{R})$ for
one-dimensional quantum systems by $N \times N$ matrices---operators
in the Hilbert space $l^2 (Z_N)$ of finite-dimensional quantum
mechanics. To this end an auxiliary factor $$\eta_N = \sqrt{\frac{2
\pi}{N}}$$ is introduced. We have found that for our purpose of
approximating the Feynman path integral it is still necessary to
introduce two additional \textit{dimensional quantities}:
\textit{length unit} $a$ and the corresponding \textit{unit of
linear momentum} $\hbar /a$. Then the position operator is
approximated by the multiplication operator in position
representation
 \begin{equation*}
  q_N\ket{\rho} = a \eta_N \rho\ket{\rho}.
\end{equation*}
For the momentum operator $p_N$ J. Schwinger had the real insight to
define it as the discrete Fourier transform of $q_N$, implying that
$p_N$ has not the form of the generally used difference operator.
Thus the momentum operator is approximated by the multiplication
operator in momentum representation
 \begin{equation*}
  p_N\ket{k} = \frac{\hbar}{a} \eta_N k\ket{k}.
\end{equation*}

Schwinger's geometric idea was to identify $Z_N$ with a
\textit{grid} in $\mathbb{R}$. For $N$ \textit{odd}, he defined a
sequence of grids $  L_N = \{a \eta_N \rho \vert \rho = 0, \pm 1,
\dots, \pm (N-1)/2  \}$. In the limit $N \rightarrow \infty$ the
grids are becoming denser and at the same time extending to the
whole real line. The grid serves to embed the finite-dimensional
Hilbert space $l^2 (Z_N)$ isometrically in $L^2 (\mathbb{R},dq)$ by
the map
 \begin{equation*}
\mathcal{I}: \ket{\rho} \mapsto \phi_{\rho} (q)=
a^{-\frac{1}{2}}\eta_N^{-\frac{1}{2}}
 \chi_{[a \eta_N (\rho - \frac{1}{2}),a \eta_N (\rho +
 \frac{1}{2})]}(q),
\end{equation*}
where $\chi_S$ denotes the characteristic function of a subset
$S\subset \mathbb{R}$. The position eigenvectors $\ket{\rho}$ are
thus mapped onto narrow normalized wave functions $\phi_{\rho} (q)$
on the real line, centered at the grid points and contracting in the
limit $N \rightarrow \infty$. Under the map $\mathcal{I}$ the
normalizations of the wave functions
$\ket{\psi}=\sum_{\rho}\psi_{\rho}\ket{\rho}$ and
$(\mathcal{I}\psi)(q)$ are related by
 \begin{equation*}
\vert \psi_\rho \vert^2 =a\eta_N \vert (\mathcal{I}\psi)(a\eta_N
\rho)\vert^2.
\end{equation*}

\section{Finite-dimensional analogue of quantum free particle}
Finite-dimensional analogue of a quantum free particle was
formulated in \cite{StovTolar} as a discrete Galilean evolution
along a finite closed linear chain. The single-step unitary time
evolution operator $C_N$ proposed there is diagonal in momentum
representation
  \begin{equation*}
  \bra{j}C_N\ket{k}= \delta_{jk} \omega_N^{-k^2}.
\end{equation*}
Transformation to position representation gives
  \begin{equation*}
 (C_N)_{\rho \sigma}= \sum_{jk}\braket{\rho}{j}\bra{j}C_N
 \ket{k}\braket{k}{\sigma}  =
 \frac{1}{N}\sum_{j=0}^{N-1}\omega_N^{-j^2 + (\rho -\sigma)j}.
\end{equation*}
The unitary operator $C_N$ fulfils relations
 \begin{equation*}
 C_N^{-1}Q_N C_N=\omega_N Q_N P_N^2, \quad C_N^{-1}P_N C_N =P_N.
\end{equation*}

Looking at free evolution in continuous phase space, we arrive to
the conclusion that operator $C_N$ should be slightly modified.
However, first consider the usual one-parameter group of unitary
operators
 \begin{equation*}
T(t)= \exp({-\frac{i}{\hbar}\frac{\hat{p}^2}{2m}t}),\quad
t\in\mathbb{R},
\end{equation*}
describing quantum evolution of a non-relativistic free particle of
mass $m$ on the real line. The corresponding $N\times N$
approximation is
 \begin{equation*}
T_{N}(\tau)=
\exp({-\frac{i}{\hbar}\frac{p_N^2}{2m}\tau\varepsilon}), \quad \tau
\in \mathbb{Z},
\end{equation*}
where we have introduced a time unit $\varepsilon$, since
dynamically defined time intervals will play a special role. Thus
$t$ shall be restricted to integer multiples $\tau\varepsilon$,
$\tau \in \mathbb{Z}$, of $\varepsilon$. The time unit $\varepsilon$
will be chosen so that (in momentum representation)
 \begin{equation*}
T_{N}(\tau)\ket{j}=
\exp({-\frac{i}{\hbar}\frac{1}{2m}(\frac{\hbar}{a}\eta_N
j)^2\tau\varepsilon})\ket{j}= \omega_N^{-\frac{1}{2}j^2
\tau}\ket{j}, \quad \tau \in \mathbb{Z},
\end{equation*}
including an additional $1/2$ factor in the exponent.
\footnote{Non-integer powers of $\omega_N$ are understood as complex
exponentials $\omega_N^w=\exp(\frac{2\pi i}{N}w)$,
$w\in\mathbb{R}$.} Our choice is in agreement with a dynamical
relation
 \begin{equation*}
 \varepsilon = \frac{ma^2}{\hbar} \qquad  \mbox{or} \qquad
 m\frac{a}{\varepsilon} = \frac{\hbar}{a}
\end{equation*}
expressing the natural fact that a particle of momentum $\hbar /a$
traverses the distance $a$ in time $\varepsilon$. Transformation to
position representation gives
  \begin{equation*}
T_{N}(\tau)_{\rho \sigma}= \sum_{jk}\braket{\rho}{j}\bra{j}T_N(\tau)
 \ket{k}\braket{k}{\sigma}  =
 \frac{1}{N}\sum_{j=0}^{N-1}\omega_N^{-\frac{1}{2}j^2\tau+(\rho - \sigma)j}.
\end{equation*}

Now in order to justify the $1/2$ factor in the exponent recall that
the free time evolution in continuous phase space $\mathbb{R}^2$ is
described by translations along $q$ with constant velocity,
  \begin{equation*}
 q(t) = q(0) + \frac{p(0)}{m}t, \quad p(t)=p(0).
 \end{equation*}
The corresponding single-step 'translation' in quantum phase space
$\Gamma_N$,
 \begin{equation*}
 Q \mapsto QP, \qquad P \mapsto P
 \end{equation*}
is equivalent to the following $SL(2,\mathbb{Z}_N)$ transformation
       \begin{equation*}
        (1,1) =  (1,0) \left( \begin{array}{cc}
        1 & 1 \\
        0 & 1
       \end{array}  \right), \quad
               (0,1) =  (0,1) \left( \begin{array}{cc}
        1 & 1 \\
        0 & 1
       \end{array}  \right).
       \end{equation*}
Its powers form an Abelian subgroup of $SL(2,\mathbb{Z}_N)$
isomorphic to $\mathbb{Z}_N$. An easy calculation shows that it is
implemented by the unitary transformation
 $$T_N(1)\ket{j} =\omega_N^{-\frac{1}{2}j^2}\ket{j}=C_{N1}\ket{j},
 $$
which, from now on, will be denoted $C_{N1}$. The modified unitary
operator $C_{N1}$ now fulfils relations
 \begin{equation*}
   C_{N1}^{-1}Q_N C_{N1}=\omega_N^{\frac{1}{2}}Q_N P_N,
 \quad C_{N1}^{-1}P_N C_{N1} =P_N,
\end{equation*}
 \begin{equation*}
 C_{N1}^{-s} Q_N^\rho P_N^j C_{N1}^{s}=
 \omega_N^{\frac{1}{2}\rho^2 s} Q_N^\rho P_N^{j+\rho s}.
\end{equation*}

\section{$N\times N$ approximation of the Feynman path integral}

 Let $\ket{q(0),0}$ and $\ket{q(t),t}$ be the state vectors of the
initial state and of the final state of a particle on $\mathbb{R}$
at times $0$, $t$, respectively. If $S[q]$ is the classical action
functional of the particle, the evolution amplitude is according to
Feynman formally written as a path integral \cite{FeynHibbs}
\begin{equation*}
\braket{q(t),t}{q(0),0}=\int e^{\frac{i}{\hbar}S[q]}
\mathcal{D}q(t).
\end{equation*}
It is understood as a sum over all continuous paths in configuration
space. According to Feynman's principle of equivalence of
trajectories, the contribution of each path should have the same
absolute value, hence contributes to the sum only a phase factor
with the phase given by the classical action in units $\hbar$
evaluated along the path.

In quantum mechanics, the path integral is traditionally defined as
a limit via discretization based on the division of the time
interval, e.g. into $n$ intervals of equal duration $\varepsilon =
t/n$. The evolution amplitude is thus written as a multiple integral
\cite{FeynHibbs}
\begin{equation*}
 \braket{q(t),t}{q(0),0}=
\end{equation*}
\begin{equation*}
 =\int_{-\infty}^{+\infty} \dots \int_{-\infty}^{+\infty}
 \bra{q(t)}e^{-\frac{i}{\hbar}H\varepsilon}\ket{q_{n-1}}dq_{n-1}
 \dots
 dq_1 \bra{q_1}e^{-\frac{i}{\hbar}H\varepsilon}\ket{q(0)},
\end{equation*}
where $q_l = q(l\varepsilon)$ and $H$ is the Hamilton operator. Each
factor---the short-time propagator---is then identified with an
exponential of the short-time action involving an approximation of
the classical Lagrangian,
\begin{equation*}
 \bra{q_{l+1}}e^{-\frac{i}{\hbar}H\varepsilon}\ket{q_l}=\frac{1}{A}
 e^{\frac{i}{\hbar}L(q_{l+1},q_l)\varepsilon},
\end{equation*}
with normalization factor $A$. For instance, for a non-relativistic
particle of mass $m$
\begin{equation*}
 H =\frac{\hat{p}^2}{2m} + V(\hat{q}),
\end{equation*}
and one computes (via momentum representation)
\begin{equation*}
\frac{1}{2\pi\hbar}\int_{-\infty}^\infty \exp({\frac{i}{\hbar}(p_l
\frac{q_{l+1}-q_l}{\varepsilon}- \frac{p_l^2}{2m}
-V(q_l))\varepsilon})dp_l =
\end{equation*}
\begin{equation}\label{potential}
=(\frac{2\pi i \hbar \varepsilon}{m})^{-\frac{1}{2}}
\exp(\frac{i}{\hbar}(\frac{1}{2}m
(\frac{q_{l+1}-q_l}{\varepsilon})^2
 - V(q_l))\varepsilon),
\end{equation}
i.e.
\begin{equation*}
L(q_{l+1},q_l)= \frac{1}{2}m \left(\frac{q_{l+1}-q_l}{\varepsilon}
\right)^2 - V(q_l) \quad \mbox{and} \quad A=(\frac{2\pi i \hbar
\varepsilon}{m})^{\frac{1}{2}}.
\end{equation*}
A sequence of $q_l$'s for each $t_l$ shall, in the limit, define a
path of the system and each of the integrals is to be taken over the
entire range available to each $q_l$. In other words, the multiple
integral is taken over all possible paths.\footnote{Recall a
quotation from Feynman's thesis (p. 69 of \cite{Feyn}): \textit{"A
point of vagueness is the normalization factor, $A$. No rule has
been given to determine it for a given action expression. This
question is related to the difficult mathematical question as to the
conditions under which the limiting process of subdividing the time
scale, required by equations such as (68), actually converges."}}

Let us return to our analogue of a \textit{free non-relativistic
particle}. The above approach will guide us in our $N \times N$
approximation with the short-time propagator induced by the unitary
operator $C_{N1}$. In this approximation $q_{l} \approx a \eta_N
\rho_{l}$, so, for a single time step,
$\braket{q_{l+1},\varepsilon}{q_l,0}$ is approximated by
\begin{eqnarray*}
 \braket{q_{l+1},\varepsilon}{q_l,0}a\eta_N &=&
 \bra{q_{l+1}}e^{-\frac{i}{\hbar}H\varepsilon}\ket{q_l}a\eta_N =\\
 =\bra{\rho_{l+1}}C_{N1}\ket{\rho_{l}}&=&
\frac{1}{N}\sum_{j_l=0}^{N-1}\omega_N^{-\frac{1}{2}j_l^2+
(\rho_{l+1} - \rho_l)j_l}.
\end{eqnarray*}
For $\tau$ time steps
\begin{equation*}
\braket{\rho_\tau,\tau\varepsilon}{\rho_0,0}=
\sum_{\rho_1,\dots,\rho_{\tau -1}}
\bra{\rho_{\tau}}C_{N1}\ket{\rho_{\tau -1}}\dots
\bra{\rho_{1}}C_{N1}\ket{\rho_{0}} =
\end{equation*}
\begin{equation*}
=\bra{\rho_{\tau}}C_{N1}^\tau\ket{\rho_{0}}=
\frac{1}{N}\sum_{j=0}^{N-1}\omega_N^{-\frac{1}{2}j^2\tau+
(\rho_{\tau} - \rho_0)j}.
\end{equation*}

The above Gauss-like sum for a single time-step can be summed up
using C.L. Siegel's Reciprocity Formula for generalized Gauss sums
\cite{Siegel60,BerndtEvans,BerndtEvans98}
\begin{equation*}
\sum_{n=0}^{\vert c \vert -1}e^{\pi i (an^2 +bn)/c} =\sqrt{\vert
\frac{c}{a} \vert} e^{\pi i(\vert ac\vert-b^2) /
(4ac)}\sum_{n=0}^{\vert a\vert -1} e^{-\pi i(cn^2 +bn)/a}
\end{equation*}
valid for $a,b,c \in \mathbb{Z}$, $ac\neq 0$, $ac+b$ even. Putting
$a=N$ with $N$ \textit{odd}, $c=1$, $n=j_l$ and $b=-2\rho -1$ with
$\rho=\rho_{l+1}-\rho_l$ one obtains
\begin{equation*}
\frac{1}{N}\sum_{j_l=0}^{N-1}\omega_N^{-\frac{1}{2}j_l(j_l -1)+
(\rho_{l+1} - \rho_l)j_l}=
\frac{1}{\sqrt{iN}}\omega_N^{\frac{1}{2}(\rho_{l+1}-\rho_l
+\frac{1}{2})^2}.
\end{equation*}
On the basis of this formula we prefer the operator
\begin{equation*}
 C_{N2}\ket{j}=\omega_N^{-\frac{1}{2}j(j-1)}\ket{j},
\end{equation*}
for unitary single-step time evolution. It satisfies simpler
relations than $C_{N1}$,
 \begin{equation*}
   C_{N2}^{-1}Q_N C_{N2}=Q_N P_N,
 \quad C_{N2}^{-1}P_N C_{N2} =P_N,
\end{equation*}
 \begin{equation*}
 C_{N2}^{-s} Q_N^\rho P_N^j C_{N2}^{s}= Q_N^\rho P_N^{j+\rho s},
\end{equation*}
while inducing the same Abelian subgroup of translations of quantum
phase space. With operator $C_{N2}$ we compute in position
representation
\begin{equation*}
\bra{\rho_{l+1}}C_{N2}\ket{\rho_{l}}=
\frac{1}{\sqrt{iN}}\omega_N^{\frac{1}{2}(\rho_{l+1}-\rho_l
+\frac{1}{2})^2}.
\end{equation*}
This result can be interpreted as the \textit{emergence of a
dimensionless Lagrangian} $\mathcal{L}_N$,
\begin{equation*}
\bra{\rho_{l+1}}C_{N2}\ket{\rho_{l}}=
\frac{1}{\sqrt{iN}}\omega_N^{\mathcal{L}_N(\rho_{l+1},\rho_l)},
\quad \mathcal{L}_N(\rho_{l+1},\rho_l)=\frac{1}{2}(\rho_{l+1}-\rho_l
+\frac{1}{2})^2.
\end{equation*}
In order to go over to the $1+1$ space-time and obtain the
corresponding local Lagrangian $L_N$, we divide by $a\eta_N$ and
express the short-time propagator
\begin{equation*}
 \braket{q_{l+1},\varepsilon}{q_l,0}=
 \frac{1}{a\eta_N}\bra{\rho_{l+1}}C_{N2}\ket{\rho_{l}}=
\end{equation*}
 \begin{equation*}
 =\frac{1}{a\eta_N}\frac{1}{\sqrt{i N}}
\omega_N^{\frac{1}{2}(\rho_{l+1}-\rho_l +\frac{1}{2})^2}=
 (\frac{2\pi i \hbar \varepsilon}{m})^{-\frac{1}{2}}
e^{\frac{i}{\hbar}\frac{1}{2}m (\frac{q_{l+1}-q_l +
\frac{a\eta_N}{2}}{\varepsilon})^2 \varepsilon}.
\end{equation*}
This result can be rewritten
\begin{equation*}
\frac{1}{\sqrt{i N}}\omega_N^{\mathcal{L}_N(\rho_{l+1},\rho_l)}=
(\frac{2\pi i \hbar \varepsilon}{m})^{-\frac{1}{2}}
e^{\frac{i}{\hbar}L_N (q_{l+1},q_l) \varepsilon}a\eta_N,
\end{equation*}
where the phase factor appearing in the short-time propagator is
seen to define the corresponding small increment of the action $L_N
\varepsilon$ which is proportional to the local Lagrangian
\begin{equation*}
 L_N =\frac{1}{2}m (\frac{q_{l+1}-q_l +
\frac{a\eta_N}{2}}{\varepsilon})^2.
\end{equation*}
Note that in the limit $N \rightarrow \infty$, we have $\eta_N =
\sqrt{2\pi/N}\rightarrow 0$ and obtain the usual form of the
short-time propagator for the free quantum particle.

We close this section with a one-dimensional particle moving in a
potential field $V(q)$. This potential has been incorporated in the
short-time propagator (\ref{potential}). To get its $N\times N$
approximation, potential $V(q)$ is sampled only at the grid points
$q_l=a\eta_N \rho_l$, $\rho_l= -(N-1)/2,\dots,(N-1)/2$. In order to
transform $V(q_l)$ into dimensionless form it should be expressed in
the energy unit
\begin{equation*}
 \frac{1}{m}(\eta_N \frac{\hbar}{a})^2=
 \frac{2\pi}{N}\frac{\hbar}{\varepsilon}
\end{equation*}
used for transforming the kinetic energy to $j^{2}/2$,
\begin{equation*}
V(q_l)=V(a\eta_N\rho_l)= \frac{2\pi}{N}\frac{\hbar}{\varepsilon}w_l.
\end{equation*}
As a result, potential is represented by a set of $N$ dimensionless
constants $w_l$. Thus the short-time propagator obtained for a free
particle is subject only to a slight modification by constants
$w_l$:
\begin{equation*}
 \braket{q_{l+1},\varepsilon}{q_l,0}=
 \frac{1}{a\eta_N}\bra{\rho_{l+1}}C_{N2}\omega_N^{-w_l}\ket{\rho_{l}}=
\end{equation*}
 \begin{equation*}
 =\frac{1}{a\eta_N}\frac{1}{\sqrt{i N}}
\omega_N^{\frac{1}{2}(\rho_{l+1}-\rho_l +\frac{1}{2})^2 -w_l}=
\end{equation*}
\begin{equation*}
 =(\frac{2\pi i \hbar \varepsilon}{m})^{-\frac{1}{2}}
\exp\left(\frac{i}{\hbar}[\frac{1}{2}m (\frac{q_{l+1}-q_l +
\frac{a\eta_N}{2}}{\varepsilon})^2-V(q_l)] \varepsilon\right).
\end{equation*}
Also in these formulae the \textit{emergence of a local Lagrangian}
is clearly manifest.

\section{Short-time propagator and mutually unbiased bases}

Let us denote the bases composed of eigenvectors of the operators
$Q_N^j P_N^\sigma$ by $\mathcal{B}_{(j,\sigma)}$. Unitary operator
$C_{N2}$ (or $C_{N1}$) plays analogous role as operator $D_N$ in our
previous study of mutually unbiased bases \textit{for prime} $N$
\cite{SulcTolar}. There the iterations of $D_N$ generated the
\textit{maximal set of $N+1$ mutually unbiased bases}
\begin{equation*}
 \mathcal{B}_{(1,0)}  \stackrel{S_N}{\rightarrow}
 \mathcal{B}_{(0,1)} \stackrel{D_N}{\rightarrow}
\mathcal{B}_{(1,1)}\stackrel{D_N}{\rightarrow} \mathcal{B}_{(2,1)}
\stackrel{D_N}{\rightarrow} \dots
\stackrel{D_N}{\rightarrow}\mathcal{B}_{(N-1,1)}
 \end{equation*}
starting with the canonical basis $\mathcal{B}_{(1,0)}$.

\textit{If $N$ is prime}, identical reasoning as in \cite{SulcTolar}
shows that the iterations of unitary operator $C_N$ (indices 1 or 2
omitted) generate in a similar way another maximal set of $N+1$
mutually unbiased bases
\begin{equation*}
 \mathcal{B}_{(0,1)}  \stackrel{S_N^{-1}}{\rightarrow}
 \mathcal{B}_{(1,0)} \stackrel{C_N}{\rightarrow}
\mathcal{B}_{(1,1)}\stackrel{C_N}{\rightarrow} \mathcal{B}_{(1,2)}
\stackrel{C_N}{\rightarrow} \dots
\stackrel{C_N}{\rightarrow}\mathcal{B}_{(1,N-1)},
 \end{equation*}
now starting with the momentum basis $\mathcal{B}_{(0,1)}$. Thus the
composite unitary operators $C_{N}^b S_N^{-1}$, $b=0,1,\dots,N-1$
produce all the bases of the maximal set when applied to the
momentum basis.

In this paper $N$ is not restricted to primes but may take arbitrary
odd value. Notwithstanding this general situation the formulae
derived in previous section show that the bases appearing in the
short-time propagators, i.e. $\{\ket{\rho}\}= \mathcal{B}_{(1,0)}$
and $\{C_N \omega_N^{-w_l}\ket{\rho}\}= \mathcal{B}_{(1,1)}$ are
mutually unbiased. Especially, for the short-time propagator
\begin{equation}\label{MUB}
\vert\bra{\rho_{l+1}}C_N\omega_N^{-w_l}\ket{\rho_l}\vert=
 \frac{1}{\sqrt{N}}
 \end{equation}
holds and it is this constant absolute value that entails mutual
unbiasedness of the bases involved. From the physical viewpoint
\textit{the state evolving after a short time interval $\varepsilon$
carries no information about the preceding state}. The trivial
information-theoretic meaning of mutual unbiasedness therefore
consists in the fact that in each single evolution step complete
loss of information occurs. Physical information is carried only by
the phase factor whose phase is proportional to the local
Lagrangian.

\section{Conclusions}

The foregoing development of $N \times N$ approximations of the
Feynman path integral is a continuation of our previous studies of
quantum mechanics over finite configuration spaces
\cite{StovTolar,SulcTolar} and of discrete path summation
\cite{ChadzTolar}. Our approximation method was first applied to
discrete time evolution of an analogue of a quantum free
non-relativistic particle in discrete $1+1$ space-time. The powers
of the evolution operator can be interpreted as a unitary
representation in $l^2(\mathbb{Z_N})$ of an Abelian subgroup of
$SL(2,\mathbb{Z_N})$ acting on quantum phase space
$\mathbb{Z_N}\times\mathbb{Z_N}$. Generalization of the $N\times N$
approximation of the short-time propagator to a particle moving in a
potential turned out to be straightforward.

Summarizing, we have shown explicitly how a Lagrangian arises as a
local phase in the short-time propagator in the $N\times N$
approximation, thus confirming G. Svetlichny's conjecture. Our paper
also brings definite answers to the questions quoted in the
Introduction for the case of a \textit{one-dimensional
non-relativistic particle moving in a potential field}.
\begin{enumerate}
\item Our results concern unitary groups $U(t)$ in $L^2(\mathbb{R})$
connected with the time evolution of a one-dimensional
non-relativistic quantum particle moving in a potential field.
Equation (\ref{MUB}) shows that position bases involved in the
$N\times N$ approximation of the short-time propagator for arbitrary
odd $N$ are mutually unbiased. In this sense also in the limit
$N\rightarrow \infty$ position bases at times $0$ and $t$ do tend to
mutual unbiasedness as $t\rightarrow 0$.
\item Our discrete analogue of quantum evolution of a free particle
in a finite-dimensional Hilbert space which accurately simulates the
Galilean evolution of a free particle on the real line was
consistently employed throughout the paper.
\item
In our paper also the secret of the prefactor $A^{-1}$ in the
short-time propagator is unveiled. Its dimension must be inverse
length in order to compensate the integration over $q$ in the
Feynman path integral. Hence in the $N\times N$ approximation it
must be related to our unit of length $a\eta_N$. Due to $\varepsilon
= ma^{2}/\hbar$,
 \begin{equation*}
\frac{a\eta_N}{A} = a \sqrt{\frac{2\pi}{N}}\sqrt{\frac{m}{2\pi
i\hbar\varepsilon}} = \frac{1}{\sqrt{iN}}
\end{equation*}
holds for single-step time evolution. This relation shows the
trivial information-theoretic meaning of $A^{-1}$: \textit{in the
dimensionless expression it is a constant corresponding to complete
loss of information in each single time step}.
\end{enumerate}

In connection with the $N\times N$ approximation we would like to
point out the suggestion of the approximate solution of the
continuous Schr\"odinger equation. Namely, Digernes, Husstad and
Varadarajan \cite{Digernes} proved a convergence theorem on
approximation of continuous Weyl systems by $N \times N$ Weyl
operators $Q_N^j P_N^\sigma$. Further, Digernes, Varadarajan and
Varadhan \cite{DigernesVV} proved a strong theorem on convergence of
eigenvalues and eigenfunctions of $N \times N$ Hamiltonians to
solutions of one-dimensional Schr\"odinger equation for potentials
satisfying $V \rightarrow +\infty$ as $\vert q \vert \rightarrow
\infty$, hence possessing discrete spectrum. This theorem provides a
justification for the approximate solution of the continuous
Schr\"odinger equation. Numerical calculations showed that the
approximation is unexpectedly good even for relatively small values
of $N$. The generalization of these results to the case of mixed
spectrum remains open. Let us note that in quantum optics, discrete
phase space $\mathbb{Z}_{n}\times \mathbb{Z}_{n}$ is employed in the
discrete approximation of the quantum phase and the conjugate number
operator \cite{PeggBarnett}.

\section*{Acknowledgements}
Partial support by the Ministry of Education of Czech Republic
(projects MSM6840770039 and LC06002) is gratefully acknowledged.

\end{document}